\documentclass[aps,pra,reprint,superscriptaddress]{revtex4-1}

\usepackage{graphicx}
\usepackage{color}
\usepackage{amsmath,amssymb}
\usepackage{bm}
\usepackage{upgreek}
\usepackage{xspace}
\usepackage{units}
\usepackage[colorlinks,urlcolor=blue,citecolor=blue,linkcolor=blue]{hyperref}
\graphicspath{{figures/}}

\newcommand{\Rb}{$^{87}$Rb\xspace}

\newcommand{\vls}{\text{vls}}
\newcommand{\thetaQWP}{\ensuremath{\theta_{\lambda/4}}\xspace}
\newcommand{\thetaN}{\ensuremath{\theta_{\text{N}}}\xspace}

\newcommand{\reffig}[1]{\mbox{Fig.~\ref{#1}}}
\newcommand{\refeq}[1]{\mbox{Eq.~(\ref{#1})}}

\newcommand{\partialD}[2]{\frac{\partial #1}{\partial #2}}

\newcommand{\abs}[1]{\vert #1 \vert \xspace}

\newcommand{\ket}[1]{\vert #1 \rangle \xspace}

\newcommand{\vect}[1]{\boldsymbol{#1}\xspace}
\newcommand{\uvect}[1]{\hat{\boldsymbol{#1}}\xspace}
\newcommand{\epvec}{\hat{\boldsymbol{\epsilon}}\xspace}
\newcommand{\HzWcm}{\text{Hz}\,\text{W}^{-1}\,\text{cm}^{2}\xspace}

\catcode`_=12
\begingroup\lccode`~=`_\lowercase{\endgroup\let~\sb}

\begin{document}

\title{Measurement and extinction of vector light shifts using interferometry of spinor condensates}

\author{A.\,A.~Wood}
\affiliation{School of Physics \& Astronomy, Monash University, Victoria 3800, Australia.}
\affiliation{School of Physics, University of Melbourne, Victoria 3010, Australia.}
\author{L.\,D.~Turner}
\author{R.\,P.~Anderson}

\affiliation{School of Physics \& Astronomy, Monash University, Victoria 3800, Australia.}

\date{\today}

\begin{abstract}
We use differential Ramsey interferometry of ultracold atoms to characterize the vector light shift (VLS) from a far-off resonance optical dipole trap at $\lambda = \unit[1064]{nm}$.
The VLS manifests as a `fictitious' magnetic field, which we perceive as a change in the Larmor frequency of two spinor condensates exposed to different intensities of elliptically polarized light.
We use our measurement scheme to diagnose the light-induced magnetic field and suppress it to $2.1(8)\times10^{-4}$ of its maximum value, by making the trapping light linearly polarized with a quarter-wave plate in each beam. 
Our sensitive measurement of the VLS-induced field demonstrates high-precision, in-vacuo interferometric polarimetry of dipole trapping light and can be adapted to measure vector shifts from other lasers, advancing the application of optically trapped atoms to precision metrology.
\end{abstract}

\maketitle

\section{Introduction}
\label{sec:introduction}
Light shifts, also known as ac Stark shifts, are the alteration of atomic energy levels due to oscillatory electric fields.
Initially a spectroscopic curiosity~\cite{autler_stark_1955}, with the advent of tunable lasers light shifts found applications ranging from Sisyphys cooling~\cite{cohen-tannoudji_nobel_1998}, to optical memories~\cite{chaneliere_light-shift-modulated_2015}, quantum information processing~\cite{deutsch_quantum-state_1998, lee_sublattice_2007, lundblad_field-sensitive_2009} and molecular physics~\cite{clark_quantum_2015}.
Light shifts and precision measurements share a tense relationship: light shifts make possible dipole traps and optical lattices and are concomitant to dispersive optical probes, all critical components in the modern metrology toolbox, but frequently also perturb the measurement, often creating undesired spatial and temporal variation.
Optical lattice clocks rely on the scalar light shift to hold and isolate atoms, but unless the light shift is made identical for both excited and ground states of the clock transition~\cite{katori_ultrastable_2003}, the timekeeping varies with the lattice intensity.
Similarly, atomic magnetometers suffer both vector and tensor light shifts~\cite{budker_optical_2013}, with the vector light shift (VLS) creating a fictitious magnetic field parallel to the probe laser wavevector.
As a result much effort is currently focused on the control, and elimination, of light shifts, not only in precision measurement but also in quantum information experiments~\cite{yu_suppressing_2013}.
The scalar light shift of a ground state vanishes at the `magic-zero wavelength'~\cite{leonard_high_2015}; lasers tuned there exert no dipole force on the atomic center of mass, but retain the vector component of the electric dipole interaction needed for synthetic gauge fields~\cite{lin_synthetic_2009}, spin-orbit coupling~\cite{lin_spin-orbit-coupled_2011} and localized spin rotations~\cite{marti_coherent_2014}.
Here we consider the inverse: measuring, and ultimately eliminating, the \emph{vector} light shift in the presence of the strong scalar light shift of an optical dipole trap (ODT).
Using differential Ramsey interferometry~\cite{wood_magnetic_2015} we measure the difference in total magnetic field experienced by two proximal spinor Bose-Einstein condensates (BECs) subject to controlled differences in intensity, making an interferometric measurement of the vector light shift experienced by the atoms.

The dipole trap unlocks the spin degree of freedom for trapped atoms, providing a spin-state independent trapping potential for studying the myriad dynamics and measurement prospects of ultracold and condensed spinor gases~\cite{stamper-kurn_spinor_2013}.
The long coherence times of ultracold atoms and small, flexible trapping volumes available in optical traps allows for the realization of precise, microscale magnetometers~\cite{budker_optical_2013}.
The spin-independence of the dipole trap for magnetometry is paramount, all trapped states ideally experience exactly the same trapping potential and systematics in the measurement from trapping potential shifts are eliminated.

In the simplest case of an ODT formed from a paraxial, single-frequency, focused far-off-resonant laser beam, Zeeman state-independent trapping is ensured if the trapping light is purely linearly polarized.
Any ellipticity in the polarization of the trapping light results in a magnetic spin-state dependent VLS in addition to the scalar light shift that forms the confining potential~\cite{mathur_light_1968,cohen-tannoudji_experimental_1972, corwin_spin-polarized_1999}.
The VLS is therefore regarded as a effective, or `fictitious', magnetic field~\cite{flambaum_magic_2008,park_measurement_2001} that vexes the inference of magnetic fields using optically trapped atoms.

Ensuring a truly spin-independent trapping potential is of great utility to many experiments, such as those seeking to measure a permanent electric dipole moment of the electron~\cite{romalis_zeeman_1999,zhu_absolute_2013}.
The VLS from optical pump and probe beams is also a concern for optical magnetometry, where pump-probe intensity fluctuations give rise to noise of the effective VLS magnetic field~\cite{budker_optical_2007}.
Precisely accounting for the VLS has important consequences for improving measurements of fundamental atomic properties, including `magic' wavelengths~\cite{leonard_high_2015}.d
Experiments where single atoms are confined in optical tweezers also contend with the VLS as a source of experimental difficulty~\cite{kaufman_cooling_2012, thompson_coherence_2013}.

For optically trapped atomic samples, the focused trapping beams makes any VLS spatially inhomogeneous and creates an effective field gradient.
In experiments with localized atoms, such as traps, released traps and beams, gradients lead to separation of the Zeeman components, known as the optical Stern-Gerlach effect~\cite{park_optical_2002}; this is frequently undesirable.
Even in vapor cell experiments where Stern-Gerlach forces are less likely to be a concern, gradients lead to dephasing of transverse magnetization and so to decay of free induction signals, or equivalently, loss of contrast in magnetic interferometry. 
Experiments that seek to explore collective spin evolution over long timescales are particularly sensitive to gradients; recent examples include magnon propagation~\cite{marti_coherent_2014} and exotic many-body states such as fragmented condensates~\cite{ho_fragmented_2000}.
These `differential light shifts'~\cite{chicireanu_differential_2011, kim_magic_2013} are a distinct phenomenon.

In Ref.~\cite{wood_magnetic_2015} we presented a magnetic gradiometer that measures the difference in Larmor frequency between a pair of BECs using differential Ramsey interferometry.
In those experiments, the trapping light was extinguished to prevent the VLS contaminating magnetic gradient measurements, limiting the maximum interrogation time and hence precision.
For this work, we use differential Ramsey interferometery of trapped atoms to measure and extinguish the VLS induced by a crossed-beam optical dipole trap.
Section~\ref{sec:theory} of this paper introduces the relevant theory that describes the VLS for our system, Section~\ref{sec:differential_ramsey} describes our technique of differential Ramsey interferometry, which we use to measure the VLS.
Section~\ref{sec:apparatus} gives an overview of the experimental apparatus and technique.
Section~\ref{sec:delayed_drop} describes experiments where we measure the VLS difference between two BECs at different positions within each dipole beam, which allows us to estimate the effective VLS gradient.
In Section~\ref{sec:in_trap}, we measure the difference in the overall VLS between two trapped BECs as a function of beam polarization and differential trap beam intensity, which allows for interrogation times of up to $\unit[15]{ms}$.
In Section~\ref{sec:spin_mixing} we show the detrimental effect of a large trap-induced VLS on an evolving spinor condensate can be entirely removed.

\section{Vector light shifts in optical dipole traps}
\label{sec:theory}
In this Section the atomic vector polarizability is described using the formalism in Refs.~\cite{rosenbusch_ac_2009,kien_dynamical_2013}.
The decomposition of the electric dipole operator into scalar, vector, and tensor components allows one to identify a `fictitious' magnetic field arising from the induced dipole moment of the atom.
In many experiments, it is the vector sum of the background magnetic field and the fictitious field $\vect{B}_0+\vect{B}_\vls$ whose direction defines the spin quantization axis, and whose magnitude determines the atomic Zeeman splitting, which we detect small variations of in this work.
In dynamic polarizability calculations of the ac Stark shift, the orientation of these vectors is often obscured by incorporating \emph{ad hoc} geometric factors.
We instead premise our analysis on the total effective magnetic field experienced by irradiated atoms, a more natural way to quantify the error in geometric approximations of the vector sum.
These approximations are readily distinguished from other trigonometric functions in expressions for the light shift and differences thereof.      

An atom interacting with an electric field \mbox{$\vect{E} = \frac{1}{2} E_0(\epvec e^{-i \omega t}+\epvec^{\ast}e^{i \omega t})$} with angular frequency $\omega$ experiences an effective, or `fictitious', magnetic field
\begin{align}
    \vect{B}_{\vls} 
    &= \frac{|E_0|^2}{4} \frac{\alpha_{nJF}^v}{2 \mu_B g_F F} (i\epvec^{\ast}\times\epvec)\, ,
\label{eq:Bvls}
\end{align}
where 
$g_F$ is the Land\'{e} $g$-factor, $\mu_B$ is the Bohr magneton, 
$\epvec$ is the polarization vector,
and $\alpha_{nJF}^v$ is the reduced vector polarizability, which depends on the atomic species, quantum state $nJF$ and the optical frequency $\omega$~\cite{kien_dynamical_2013}.

A large body of theoretical work exists devoted to \emph{ab-initio} calculation of atomic polarizabilities for alkali metal atoms, including scalar and tensor polarizabilities at the important dipole trapping wavelength $\lambda=\unit[1064]{nm}$~\cite{arora_state-insensitive_2012}.
However while Ref.~\cite{arora_state-insensitive_2012} reports all tensor components (i.e. scalar, vector and rank-2 tensor) of the polarizability at $\unit[770]{nm}$, to our knowledge the vector polarizability of \Rb at $\lambda=\unit[1064]{nm}$ has not been reported.
In the limit of large detunings compared to the hyperfine splittings~\footnote{
    A treatment accounting for the hyperfine splitting (see Eqs.~B.36, B.37 of~\cite{kien_dynamical_2013}) differs from Eqs.~(\ref{eq:reduced_nJF},\ref{eq:polarizability_nJ}) by \unit[200]{ppm} at $\lambda=\unit[1064]{nm}$.
    As this is less than other contributions to the vector polarizability (including transitions to higher principal quantum numbers and core electron corrections) and the precision reported here, a fine-structure treatment suffices.
    },
the vector polarizability is given by
\begin{align}
\label{eq:reduced_nJF}
    \alpha^v_{nJF} &= (-1)^{J+I+F}\sqrt{\frac{2F(2F+1)}{F+1}}
    \left\{
    \begin{array}{ccc}
        F & 1 & F \\  
        J & I & J \\
    \end{array}
    \right\}
    \alpha^{(1)}_{nJ} \, ,
\end{align}
and
\begin{align}
\label{eq:polarizability_nJ}
    \alpha^{(1)}_{nJ} = &(-1)^{J}\sqrt{3}\,\sum_{n', J'}(-1)^{J'}
    \left\{
    \begin{array}{ccc}
        1 & 1 & 1 \\  
        J & J' & J
    \end{array}
    \right\}
    \left|\left(n'J' \left\|\vect{d}\right\| n J \right)\right|^2 \notag \\
     & \times \frac{1}{\hbar}\text{Re}\left(
    \frac{1}{\omega_{n'J'n J}-\omega-i\Gamma_{n'J'n J}/2}\notag\right.\\
    & \hspace{6em}- \left.\frac{1}{\omega_{n'J'n J}+\omega+i\Gamma_{n'J'n J}/2}
    \right),
\end{align}
where $\omega_{n' J' n J} = \omega_{n' J'}-\omega_{n J}$ is the resonant frequency of the $\ket{nJ} \rightarrow \ket{n'J'}$ fine-structure transition, $\Gamma_{n' J' n J}$ is its natural linewidth, $\left\{J\, F \dots \right\}$ denotes the Wigner 6$j$-symbol, and $\left( n'J' \left\|\vect{d}\right\| n J \right)$ the reduced dipole matrix elements~\footnote{Our use of the reduced elements $\left( n'J' || \vect{d} || n J \right)$ is related to the other elements used in the literature by  $\left( n'J' || \vect{d} || n J \right)  = (-1)^{J'-J} \sqrt{2J+1} {\langle n J || \vect{d} || n' J' \rangle}$~\cite{steck_rubidium_2010}, but some authors do not follow this convention~\cite{kien_dynamical_2013,arora_state-insensitive_2012}.}.
For our purposes it is sufficient to consider the contributions from only the D1 ($J'=\tfrac{1}{2}$) and D2 ($J'= \tfrac{3}{2}$) lines, the dipole matrix elements of which are tabulated in Ref.~\cite{steck_rubidium_2010}.
We calculate the vector polarizability to be $\alpha ^v_{nJF}=\unit[2.366\times10^{-40}]{C\,m^2/V}$ for the \Rb $F = 1$ hyperfine level of the $5\,^{2}\text{S}_{1/2}$ ground state at $\lambda = \unit[1064]{nm}$.
Polarizability units sadly remain multifarious; $\alpha^v_{nJF}=\unit[2.126\times 10^{-24}]{cm^3}$ in cgs units and $\alpha^v_{nJF}=\unit[14.35]{a_0^3}$ in atomic units, with $a_0$ the Bohr radius.
Uncertainty in the theoretical calculation is dominated by neglect of the blue and ultraviolet lines, and are well below the experimental uncertainty of our result below.  
\begin{figure}
    \centering
    \includegraphics[width=\columnwidth]{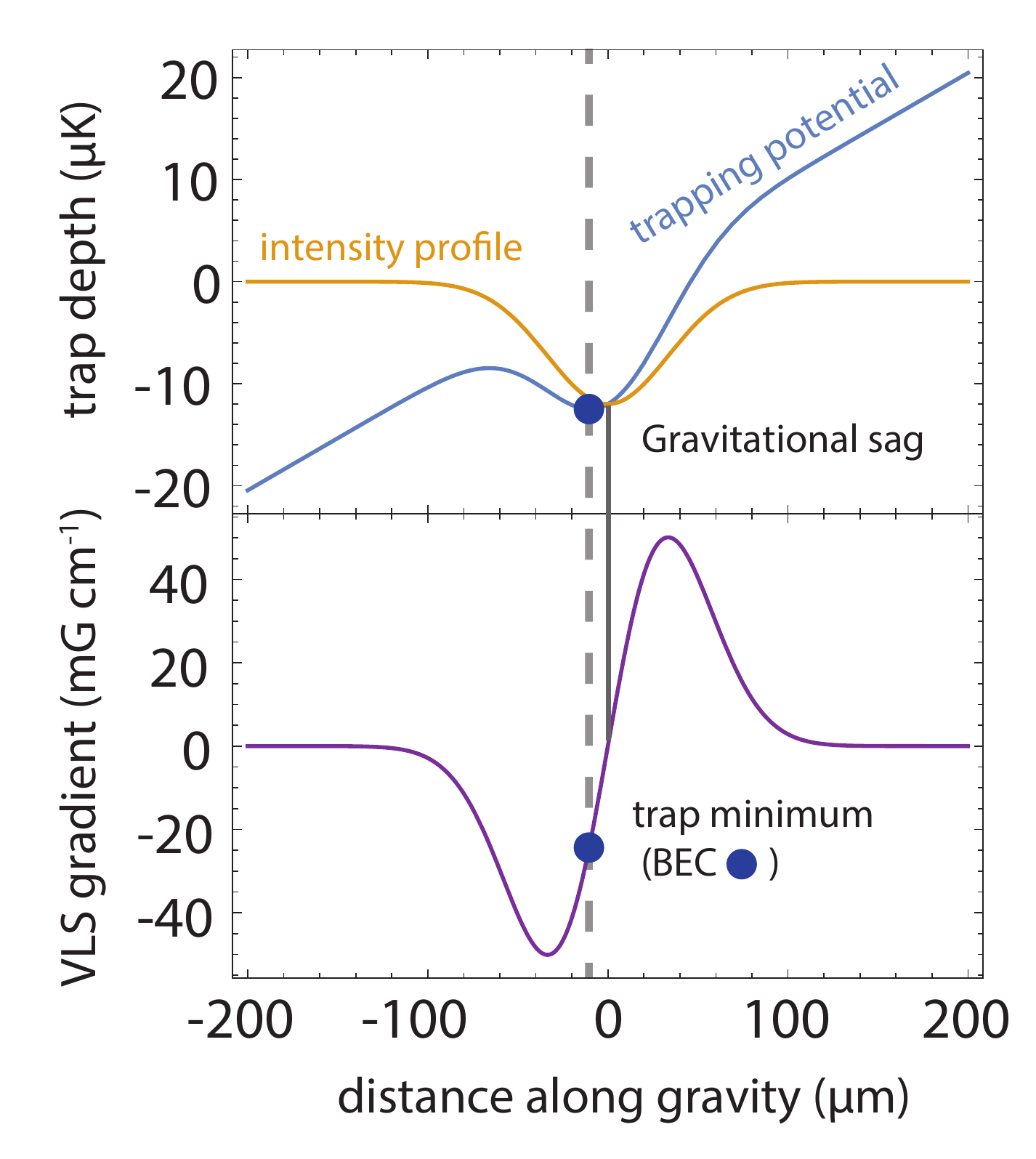}
    \caption{
    \label{fig:vls_potential_ph}
        (Top panel) The scalar light shift from a focused laser beam (orange) creates a spin-independent trapping potential, proportional to the (inverted) intensity profile.
        Gravity acts to tilt and shallow the potential (blue), with a trap minimum offset from the intensity maximum.
        (Bottom panel) A BEC (circle) trapped in this potential experiences a greater intensity variation than at the beam center, accentuating the gradient of the vector light shift.
        For the beam power ($\unit[550]{mW}$) and waist ($\unit[67]{\upmu m}$) shown here -- similar to those used in our experiment -- gravity shifts the trap minimum by $\unit[10]{\upmu m}$, where the effective field magnitude is $B_\vls=\unit[0.3]{mG}$ for a polarization circularity of $\mathcal{C} = 0.07$.
        The effective field gradient at the trap minimum is $\partial B_\vls / \partial z = \unit[24]{mG/cm}$ due to the tightly focused beam.
    }
\end{figure}

The light polarization can be expressed in terms of the right and left handed circular polarization unit vectors $\epvec_R$ and $\epvec_L$ as
\begin{equation}
\label{eq:polarization_vector}
    \epvec = \sin\left(\theta+\frac{\pi}{4}\right) \epvec_L + e^{2 i\phi} \cos\left(\theta+\frac{\pi}{4}\right) \epvec_R \, .
\end{equation}
The angle $\theta$ characterizes the degree of circular polarization; $\mathcal{C} =|\epvec^{\ast}_L \cdot \epvec|^2 - |\epvec^{\ast}_R \cdot \epvec |^2 = \sin 2\theta$ is the circularity, equal to the normalized Stokes parameter $S_3/S_0$ ($\theta = \mathcal{C} = 0$ for linearly polarized light).
$\phi$ is the orientation of the polarization axis (modulo $\pi$).
A quarter-wave plate can be used to adjust the circularity, e.g. horizontally polarized light ($\epvec = \uvect{x}$) that has passed through a quarter wave plate oriented at an angle $\theta$ counter-clockwise from vertical has the same circularity as the state in \refeq{eq:polarization_vector}.
For plane waves, the cross product in \refeq{eq:Bvls} is thus $i \epvec^{\ast} \times \epvec = - \mathcal{C} \uvect{k} = - \sin 2\theta\,\uvect{k}$, where $\uvect{k}$ is the unit wavevector of the plane wave.
Thus the VLS effective field $\vect{B}_{\vls}$ induced by a beam is always along the propagation direction of the beam. 

The vector polarizability induces a change in the energy of the Zeeman states $\ket{F,m_F}$ by an amount referred to as the vector light shift 
\begin{align}
\label{eq:Evls_defn}
    \Delta E_\vls &= \mu_B g_F m_F \left(\abs{\vect{B}_0+\vect{B}_\vls} - \abs{\vect{B}_0} \right) \notag \\
                  &= \mu_B g_F m_F B_\vls \left( \cos \nu + \frac{\sin^2\nu}{2} \frac{B_\vls}{B_0} + \mathcal{O}\left( \frac{B_\vls}{B_0} \right)^2 \right) \notag \\
                  &\approx - I_0 \left(\frac{\alpha^v_{nJF}}{4 c \epsilon_0}\right) \sin 2\theta \, \cos\nu \, \frac{m_F}{F} \, ,
\end{align}
where $\nu$ is the angle between $\vect{B}_0$ and $\vect{B}_\vls$, $B$ denotes field magnitudes, and $I_0 = (c\,\epsilon_0\left/2)| E_0 \right|^2$ is the optical intensity.
In the final line of \refeq{eq:Evls_defn}, only the first term of the series expansion in $B_\vls/B_0$ is retained, as per convention when expressing the vector light shift~\cite{derevianko_theory_2010}, valid when $B_\vls \ll B_0$ or $\sin^2 \nu \ll 1$.
Consequently the Zeeman shift due to $\vect{B}_{\vls}$ is maximized when the magnetic bias field is parallel or anti-parallel with the beam wavevector, i.e. $\cos \nu = \pm 1$ in \refeq{eq:Evls_defn}.
Conversely, a magnetic bias much stronger than $B_\vls$ perpendicular to the beam propagation is one approach to minimizing the VLS, but this option is not always available to the experimenter.

Spatially inhomogeneous laser intensity results in a spatially inhomogeneous VLS.
For a trapped cloud at the intensity maximum of a Gaussian beam, the intensity and hence VLS is locally quadratic, and so for a cloud much smaller than the beam waist, is approximately homogeneous.
Gravitational sag of the dipole potential results in displacement of the trap minimum from the intensity maximum, and thus trapped atoms experience a gradient of the VLS along the direction of gravity.
Due to the tight focusing of trapping beams this gradient can be of order $\unit[10]{mG\,cm^{-1}}$ for common trap intensities, even when reasonable care has been taken to make residual elliptical polarization quite small; \reffig{fig:vls_potential_ph} shows this effect.
Experiments requiring coherent evolution over long timescales, where even small dephasing effects from magnetic gradients are problematic, demand gradients at least an order of magnitude smaller than this, requiring elimination of the inhomogeneous VLS.

\begin{figure*}[ht]
    \centering
    \includegraphics[width=\textwidth]{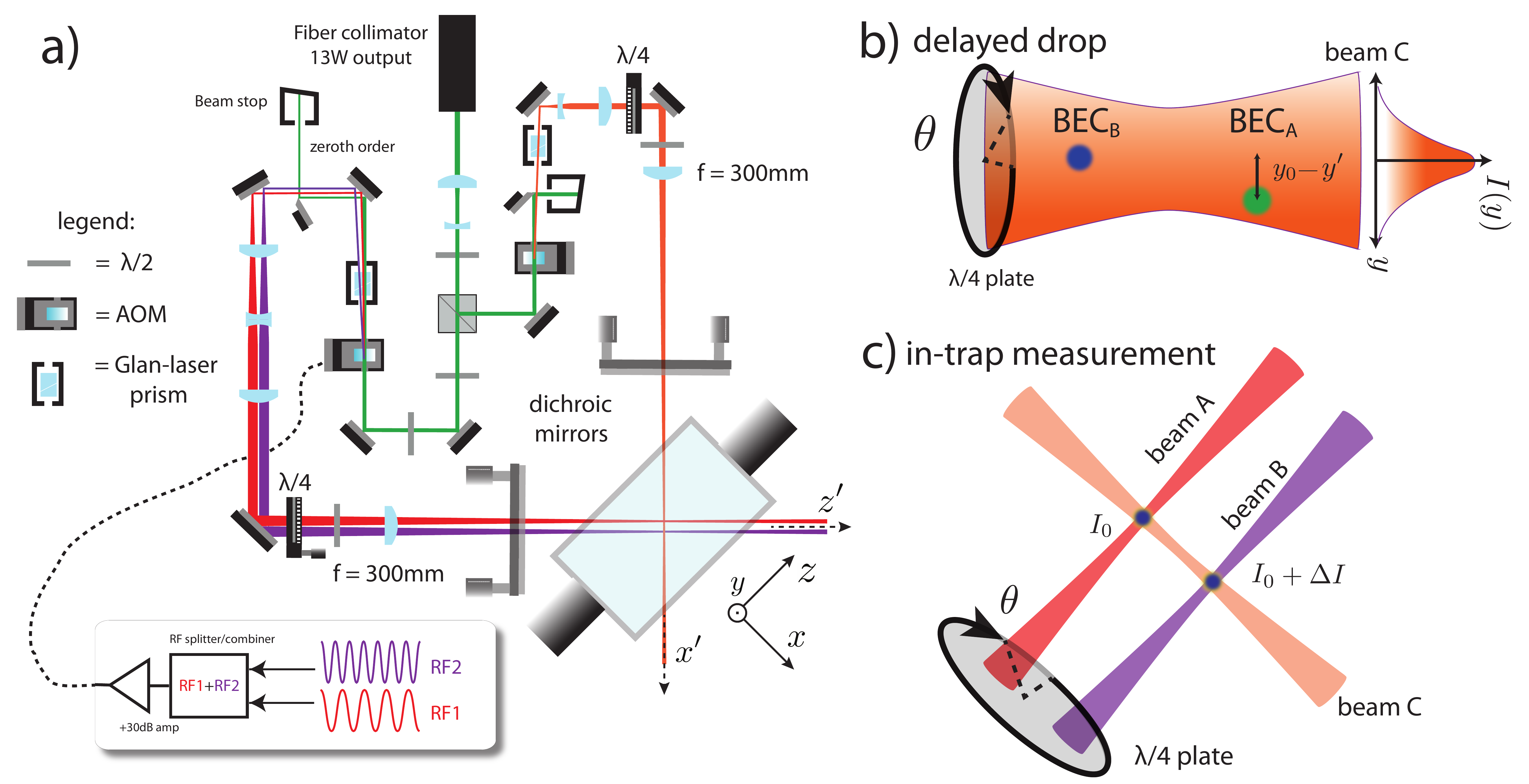}
    \caption{
        Schematic of the experimental apparatus (a) and different configurations to measure vector light shifts (b--c).
        The position and amplitude of each dipole trapping beam is controlled by an AOM; introducing a second rf frequency to an AOM allows independent control of two dipole beams propagating along the same direction.
        Splitting either the $x'$- or $z'$-oriented dipole beams into two beams $A$ and $B$ allows us to position two BECs along the axial extent of a third, orthogonal beam $C$.
        We measure the difference in VLS-induced fields at the location of each condensate using two distinct methods:
        (b) \emph{Delayed drop method.} By turning beams $A$ and $B$ off at different times, each BEC falls a different distance prior to the interrogation, sampling a different intensity of beam $C$ alone.
        (c) \emph{In-trap interferometry.} Higher precision is achieved by keeping the BECs in trap.
        We vary the relative intensity of beams $A$ and $B$ (which share a common polarization), and measure the intensity dependence of the VLS-induced field.
        For each method, we vary the magnitude of the VLS-induced field and its gradient by rotating a QWP in each beam path.
    }
    \label{fig:vls_super_fig}
\end{figure*}

\section{Differential Ramsey interferometry}
\label{sec:differential_ramsey}
We perform simultaneous Ramsey interferometry on a pair of condensates to detect magnetic field difference at their respective positions.
We have previously used this technique in the absence of dipole trapping light to measure the gradient tensor of ambient magnetic fields~\cite{wood_magnetic_2015}.
In that work we circumvented the effects of vector light shifts by releasing the atoms from the ODT prior to interrogating the internal states, yet this freefall operation restricted the sensitivity and spatial resolution of the technique. 
Minimizing the vector light shift from the dipole trapping light allows magnetic gradiometry to be performed on trapped condensates, offering significant enhancement in the potential sensitivity of these devices.
Viewed as a co-magnetometer, these improvements enhance the magnetic field resolution per spatiotemporal bandwidth $\delta B^2 T \, V$; here we extend the interrogation time from $T=\unit[3]{ms}$ to $\unit[15]{ms}$ (limited by inhomogeneous dephasing due to the residual ambient field gradient) and retain a lower sensor volume $V$ (and thus a higher density) than that of a freely falling, expanding condensate~\cite{wood_magnetic_2015}.
\begin{figure}
    \centering
    \includegraphics[width=\columnwidth]{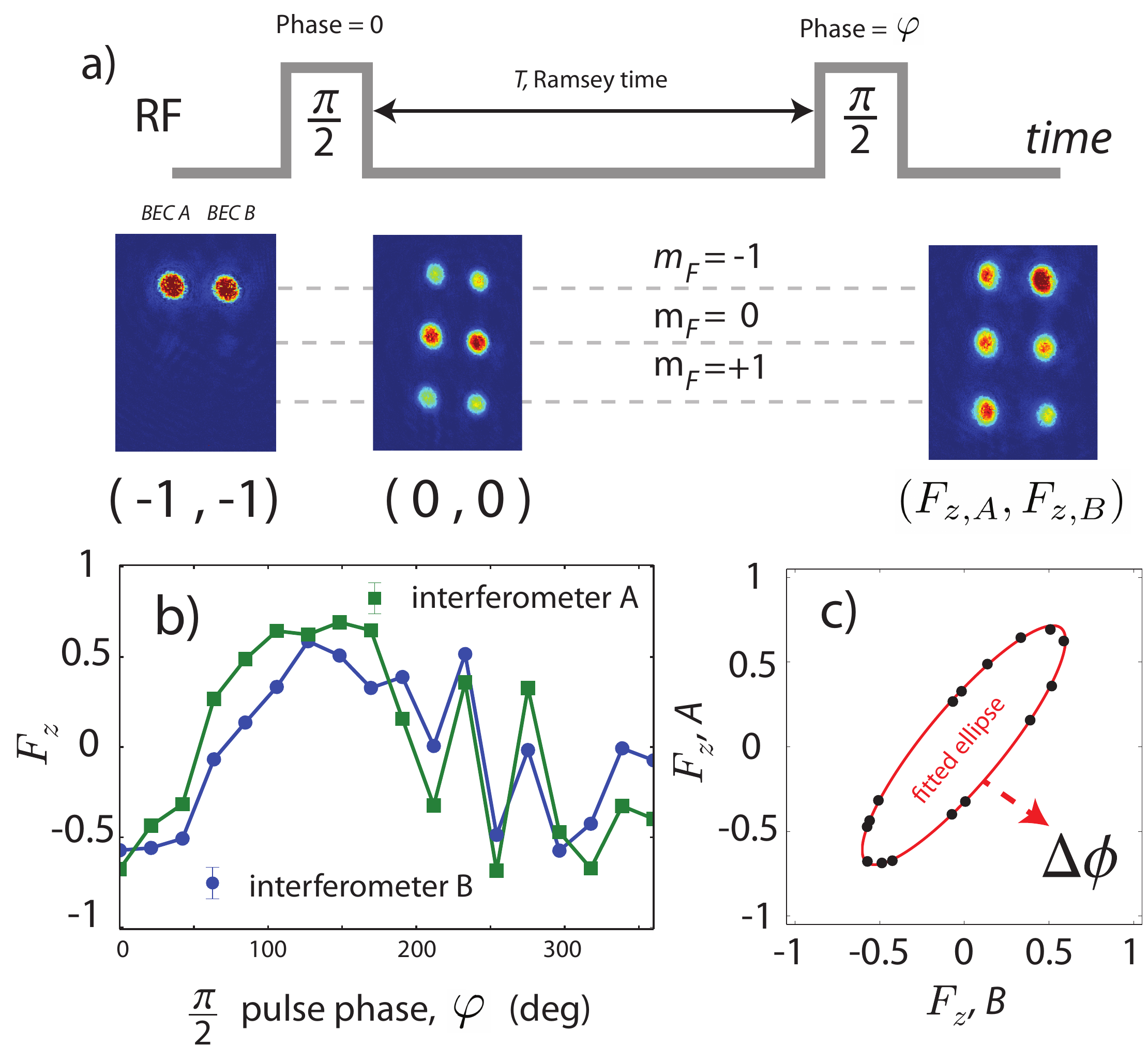}
    \caption{Differential Ramsey interferometry and elliptical data reduction.
    a) Representative Stern-Gerlach absorption images after time of flight for each stage of the Ramsey pulse sequence.
    We measure the spin projection of each condensate $F_z$; varying the second $\pi/2$-pulse phase for each iteration of the experiment traces out phase domain Ramsey fringes.
    Shot-to-shot field fluctuations scramble the phase of each interferometer, but plotting the output of one interferometer against the other yields an ellipse (c), from which we can extract the relative phase $\Delta \phi$.}
    \label{fig:ramsey_fig}
\end{figure}

We confine two proximate ultracold atom clouds in separate dipole traps, composed of two beams $A$ and $B$ intersecting a third, orthogonal beam $C$ (\reffig{fig:vls_super_fig}(c)).
Beams $A$ and $B$ (\reffig{fig:vls_super_fig}(a)) are produced by a single acousto-optic modulator (AOM) driven by two rf frequencies providing independent frequency and optical power control.
The positions of the condensates can be shifted axially along beam $C$ by changing the frequencies of the rf tones, and the intensity of each beam is controlled by adjusting the rf drive amplitudes.
The configuration can be applied to dipole trapping light propagating along either the $x'$ or $z'$ axes.
Zero-order quarter-wave and half-wave plates (QWP, HWP) are placed in the path of both $x'$ and $z'$ beams to control their polarization.

We initiate the Ramsey interferometry sequence with a radiofrequency $\pi/2$-pulse that drives transitions between the three-level system formed by the Zeeman sublevels $\ket{F=1, m_F = -1, 0, +1}$ of the \Rb $5^2\text{S}_{1/2}$ ground state.
The two condensates constitute independent spin-1 interferometers, which are separated by $10$--$\unit[100]{\upmu m}$ and thus experience the same rf Rabi frequency $\Omega_{\text{rf}}$ generated by a coil antenna several cm away.
After an interrogation time $T$, a second $\pi/2$-pulse converts the phase acquired during the interval into relative populations of the Zeeman states in each interferometer.
The phase difference between the two interferometers $\Delta \phi$ is a measure of the magnetic field difference at each condensate
\begin{equation}
\label{eq:delta_B_defn}
    \Delta B \equiv |\vect{B}(\vect{r}_A)| - |\vect{B}(\vect{r}_B)| = \frac{\Delta \phi}{\gamma T} \, ,
\end{equation}
where $\gamma$ is the gyromagnetic ratio.
Magnetic field fluctuations scramble the phase of each interferometer for interrogation times $T > \unit[1]{ms}$ in our experiments.
Simultaneous interferometry provides profound rejection of common-mode magnetic field fluctuations allowing relative phase measurements at much longer times~\cite{wood_magnetic_2015}.
The interrogation time is limited by the Zeeman coherence time of each condensate, which is set by magnetic inhomogeneity within each trap.
Reducing the VLS and thus its gradient in each trap prolongs the useful interrogation time $T$; which in the limit of vanishing VLS is ultimately set by background magnetic field gradients that must be reduced independently.
Here these are of order $10$--$\unit[20]{mG/cm}$ and we use a trapped interrogation time $T=\unit[15]{ms}$.

\section{Apparatus}
\label{sec:apparatus}
We produce \Rb BECs of $10^5$ atoms in the $\ket{F = 1, m_F = -1}$ state by loading a hybrid optical dipole-magnetic quadrupole trap~\cite{lin_rapid_2009} and transferring the atoms into the two separate optical dipole traps formed by three beams.
The dipole trapping light originates from a $\unit[20]{W}$ single-frequency $\unit[1064]{nm}$ fiber laser (Keopsys CYFL-20W-MEGA).
We are able to create approximately equal atom number BECs in each dipole potential well by loading the hybrid trap with an AOM splitting one of the beams by $\sim \unit[100]{\upmu m}$ and evaporating to BEC by reducing the intensity in all three beams equally over $\unit[5]{s}$.
The two dipole beams propagate along axes $z'$ and $x'$ with $1/e^2$ waists $\unit[67]{\upmu m}$ and $\sim \unit[100]{\upmu m}$ respectively and may be split independently, as shown in~\reffig{fig:vls_super_fig}(a).
Larger separations that clearly resolve the two condensates are obtained by smoothly increasing the AOM drive frequency difference after condensing into each trap, thus transporting the two BECs to the desired separation.

A magnetic bias field is then applied and the experiment staging is paused until the start of next cycle of the $\unit[50]{Hz}$ AC power line for improved repeatability.
We then perform the Ramsey interferometry sequence.
Ramsey fringes are obtained by varying the phase $\varphi$ of the second $\pi/2$-pulse with each iteration of the experiment.
Depending on the configuration in which the BECs are positioned in the traps (outlined in Sections~\ref{sec:delayed_drop} and~\ref{sec:in_trap}) the BECs are exposed to different intensities of elliptically polarized dipole trapping light and thus accumulate a relative phase in the Ramsey sequence proportional to the differential vector light shift.
The dipole trapping light is then completely extinguished and the atoms are absorption imaged after 23\,ms time of flight with a resonant probe laser.
A $\unit[50]{G/cm}$ Stern-Gerlach field gradient is applied for $\unit[3]{ms}$ during time of flight to spatially separate the Zeeman components of both BECs.
We compute the spin projection $F_z = \sum_{m_F}m_F N_{m_F}/\sum_{m_F} N_{m_F}$ for each interferometer $A, B$ for each applied pulse phase $\varphi$.
We are able to extract the relative phase $\Delta \phi$ in the presence of strong common-mode phase noise using an elliptical reduction of $(F_{z,A}, F_{z,B})$ data~\cite{foster_method_2002, fitzgibbon_direct_1999,szpak_guaranteed_2012,wood_magnetic_2015}.
Figure~\ref{fig:ramsey_fig} schematically shows the Ramsey pulse sequence, Stern-Gerlach absorption images, and relative phase extraction using the elliptical data reduction.

\section{Vector light shift characterization}
\label{sec:delayed_drop}
We first probe the VLS across a single dipole beam as a function of its polarization.
One dipole trapping beam is split into two beams, $A$ and $B$, which we use to prepare two BECs at different axial positions of the crossing beam (beam $C$), as shown in Figure \ref{fig:vls_super_fig}\,(b).
We then extinguish one of the split beams (beam $A$): since the crossing beam alone is insufficient to support the BEC against gravity, BEC$_A$ falls.
After a short (up to $\unit[5]{ms}$) delay, we extinguish the second split beam (beam $B$), so that the two clouds are now also \emph{vertically} separated along the radial axis of beam $C$ by the distance fallen by BEC$_A$, typically $35$-$\unit[45]{\upmu m}$.
The two clouds now sample the transverse intensity profile of beam $C$, and when beam $C$ has some ellipticity, the two BECs thus sample the differential VLS across this radial extent of the beam.
Immediately after extinguishing beam $B$ we begin the Ramsey sequence, while beam $C$ is still on.
During the Ramsey interrogation time, the two clouds continue to fall through the beam, but for short enough times ($\sim \unit[250]{\upmu s}$) this motion can be neglected.
We refer to this scheme as the `delayed-drop' technique.
\begin{figure}
    \centering
    \includegraphics[width=\columnwidth]{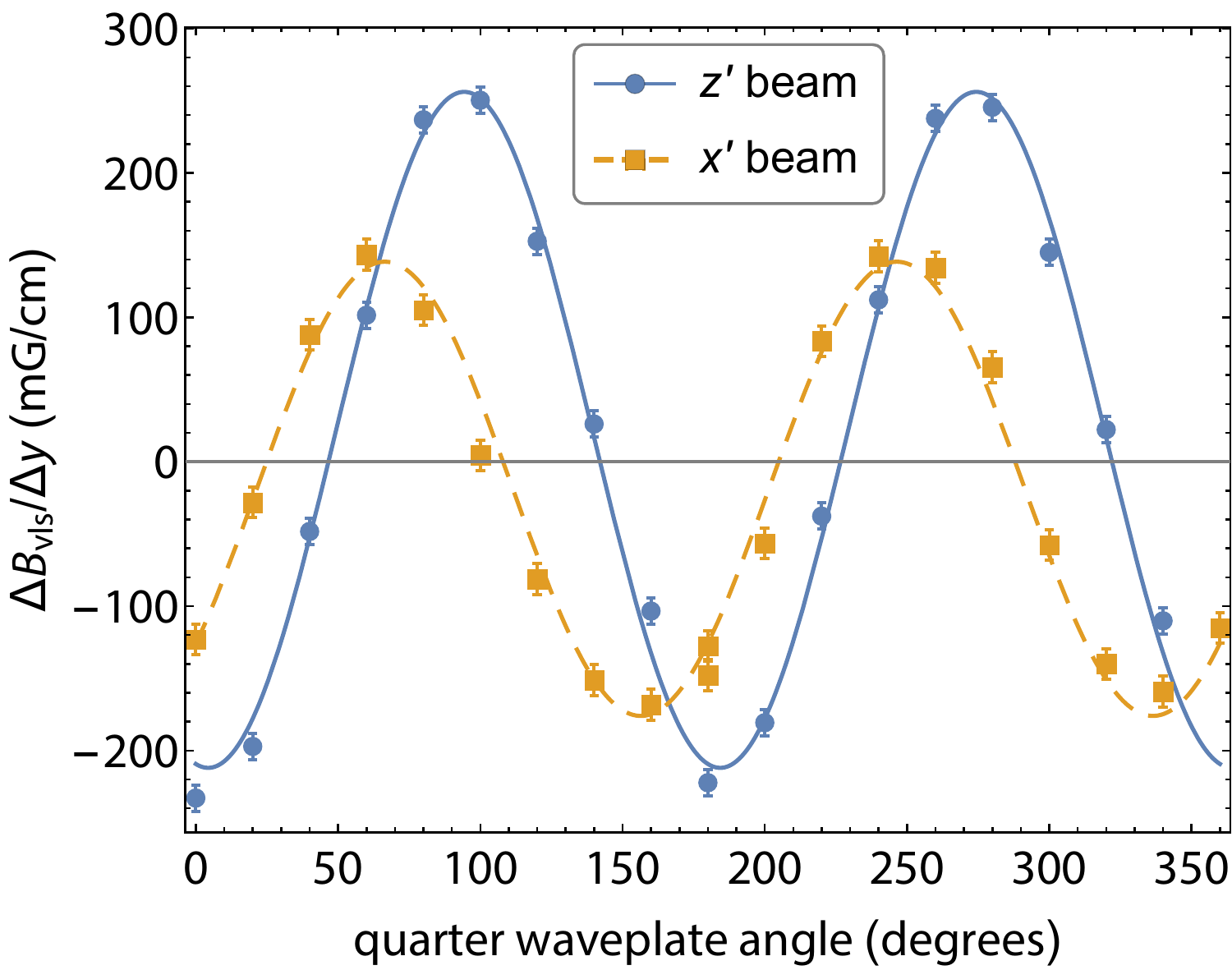}
    \caption{
        Magnetic field difference (right axis) and gradient (left axis) induced by the vector light shift from the $z'$-oriented dipole beam (circles) and $x'$-beam (squares), for variable polarization circularity.
        The VLS magnetic field difference $|\Delta B_{\vls}| < \unit[1]{mG}$ at two locations of each dipole beam (separated by $\Delta y = \unit[41.7]{\upmu m}$) are probed using differential interferometry.
        Rotating a quarter-wave plate placed in the path of each beam changes the vector light shift and thus the light-induced field gradient $\Delta B_\vls / \Delta y$.
        Biasing the measurement with three linearly independent background magnetic fields allows us to determine the direction of $\vect{B}_\vls$ for each beam ($\cos\varphi$ in \refeq{eq:delta_B_approx}).
        Solid ($z'$-beam) and dashed ($x'$-beam) lines are a phenomenological fit from which we determine the peak gradient induced by each beam (\refeq{eq:delta_B_approx}).
        }
    \label{fig:beam01}
\end{figure}

The measured relative phase depends on the difference in VLS-induced field strengths experienced by each BEC -- which is proportional to the absolute VLS of beam $C$ -- in addition to any difference in the background magnetic field.
Neglecting the axial dependence of intensity and any background magnetic field differences, the intensity difference the BECs sample by virtue of falling distances $y_A$ and $y_B$ results in a magnetic field difference that takes the form 
\begin{align}
\label{eq:delta_B_approx}
    \Delta B &= | \vect{B}_0 + \vect{B}_\vls(\vect{r}_A)| - | \vect{B}_0 + \vect{B}_\vls(\vect{r}_B)| \notag \\
             &= \underbrace{\left( B_{\vls,A} - B_{\vls,B} \right)}_{\equiv \Delta B_\vls} \left( \cos \nu + \mathcal{O}\left( \frac{B_{\vls,A} + B_{\vls,B}}{B_0} \right) \right) \, ,
\end{align}
where $B_{\vls,i} = |\vect{B}_\vls(\vect{r}_i)|$ and $\nu$ is the angle between the VLS field $\vect{B}_\vls$ and the background field $\vect{B}_0$.
The VLS field difference $\Delta B_\vls = \Delta B_{\vls,\text{max}} \sin\,2 \theta$, with $\theta$ the polarization imperfection of beam $C$.
This configuration can be implemented on either dipole beam, allowing for independent measurement of the VLS from each beam.
It is also a spatially-resolved measurement that can determine the VLS gradient as the vertical distance between the BECs is easily measured.

The glass vacuum cell and other optical components (such as the dichroic mirrors used to steer the dipole beams) exhibit birefringence, so a linearly polarized beam before these components will become elliptically polarized to some extent at the atoms~\cite{steffen_note:_2013}.
To cancel the VLS, the QWP is rotated to an angle that prepares an elliptically polarized state that, when subjected to the phase retardance of post-QWP optics, is linearly polarized in the vacuum cell.
It is possible to convert an arbitrary elliptically polarized state into a linear state using a QWP set to the appropriate angle.
To make a fully left- or right-circularly polarized state or an arbitrarily oriented linear state requires an additional HWP.
For the experiments described herein, we use a HWP of fixed orientation and rotate the QWP. 

We consider the passage of an elliptically polarized beam of light through a QWP with fast axis at angle $\theta$ and a birefringent composite phase retarder (representing the glass vacuum cell) which imparts a relative phase between linear polarization components of $\phi_k$, with a fast axis at $\theta_k$.
For linearly polarized (vertical) input light, and small birefringent phase shifts $\phi_k$, the circularity seen by the atoms can be shown using Jones calculus to be 
\begin{align}
\label{eq:circularity}
    \mathcal{C} &= -i \epvec^{\ast} \times \epvec \cdot \uvect{k} \notag \\
                &= \cos \phi_k \sin 2\theta + \sin \phi_k \cos 2\theta \sin 2(\theta-\theta_k) \, .
\end{align}
In the limit of small birefringence ($\phi_k \ll 1$) \refeq{eq:circularity} is well approximated by $\mathcal{C} = \sin 2\theta$, the circularity of the state in \refeq{eq:polarization_vector}, which we assume herein.
We note that precisely nulling the circularity for arbitrary birefringence $\phi_k$ amounts to finding the root of \refeq{eq:circularity} in $\theta$; operationally this achieved by rotating a quarter waveplate until the VLS is nulled as in Section~\ref{sec:in_trap}.

Figure~\ref{fig:beam01} shows the extent of the undesired effective field gradient induced by dipole trapping beams,
and makes plain the precision with which the dipole trap polarization must be controlled to achieve the order $\unit[1]{mG/cm}$ or lower gradients demanded by the coming generation of precision magnetometry and magnonics experiments~\cite{campbell_magnetic_2016}.
The relative phase $\Delta \phi$ is measured for different beam polarizations  by performing differential Ramsey interferometry across a $360^\circ$ range of QWP rotation angles.
The contribution to the relative phase from the background magnetic field gradient $\Delta \phi_{\text{bg}}$ is measured by extinguishing all dipole light during the Ramsey sequence, and subtracted from $\Delta \phi$ before using \refeq{eq:delta_B_defn} to impute the magnetic field difference $\Delta B$ in \refeq{eq:delta_B_approx}.
Repeating this for three linearly independent bias magnetic field directions allows us to independently determine the direction and magnitude of $\vect{B}_{\vls}$ for each beam ($\cos\nu$ and thus $\Delta B_{\vls}$ in \refeq{eq:delta_B_approx}).
We determine the effective VLS gradient $\Delta B_{\vls} / \Delta y$ plotted in \reffig{fig:beam01} as a function of QWP angle using the measured vertical displacement $\Delta y$ of each BEC in freefall.
The QWP angle(s) at which $\Delta B_{\vls}$ (and thus the VLS gradient) vanish are referred to here as the nulling angles.
The maximum VLS gradient shown in \reffig{fig:beam01} occurs when the beams are almost perfectly circularly polarized, and is $\Delta B_{\vls,\text{max}} / \Delta y = \unit[234(5)]{mG/cm}$ and $\unit[157(5)]{mG/cm}$ for the $z'$- and $x'$- beams, respectively. 

\section{Trapped-atom interferometery}
\label{sec:in_trap}
Increasing the interrogation time improves the magnetic sensitivity of differential Ramsey interferometry linearly with $T$.
In this Section we use the longer interrogation time afforded by trapped condensates for a more precise measurement and thus deeper nulling of the VLS than with atoms in freefall.
Such nulling is imperative for interrogation times beyond $\unit[1]{s}$, and should make possible magnetic sensitivities that exceed other microscopic precision magnetometers~\cite{wood_magnetic_2015}.

The measurement described in Section~\ref{sec:delayed_drop} was limited to interrogation times $T \leq \unit[5]{ms}$ as the residual dipole force provided by a single beam was insufficient to overcome gravity.
The available measurement time has been shown not to be limited by Zeeman coherence but by atom loss from the condensate~\cite{higbie_direct_2005}.
Unlike collisional effects in thermal atomic vapors, spin-changing collisions at ultracold temperatures conserve the total spin projection of the pair onto the quantization axis.
Rather than irreversible loss of interferometric contrast, ultracold collisions drive reversible dynamics of the transverse spin length on timescales even longer than the interrogation times used here.
The dominant dephasing mechanism is inhomogeneous broadening, resulting from each condensate sampling a spatially varying Zeeman shift (from ambient magnetic gradients) across its extent.

\subsection{Trapped-atom interferometry: Method}
To perform in-trap differential interferometry, we keep the two BECs in their respective dipole traps and create an intensity difference between beams $A$ and $B$, with both beams sharing a common polarization defined by a single QWP (\reffig{fig:vls_super_fig}(c)).
In addition to the $\sim \unit[4]{MHz}$ frequency difference between beams $A$ and $B$, used to control the position of each condensate, the intersecting dipole beams ($A$ and $C$, or $B$ and $C$) have a nominal relative detuning of $\unit[200]{MHz}$, which is necessary to negate any standing-wave effects resulting from their superposition.
Depending on the polarization orientation of the intersecting beams, there is either an intensity ($\text{lin}\perp\text{lin}$) or polarization ($\text{lin}\,||\,\text{lin}$) modulation~\cite{grimm_optical_2000} which moves at $2\pi\times \unit[200]{MHz}/\left|\vect{k}_{x'}-\vect{k}_{z'}\right| = \unit[152]{m\,s^{-1}}$~\cite{pethick_bose-einstein_2008}.
The atoms experience a time-average of this rapidly moving lattice; in the case of scalar light shifts this results in the total effective intensity being the sum of the two intensities.
In the case of vector light shifts, it is perhaps less obvious that the sum of polarizations from each beam, inserted into \refeq{eq:Bvls} and time-averaged, results in a \emph{vector} sum of the fictitious fields of the independent beams, e.g. $\vect{B}_{\vls} = \vect{B}_{\vls,A} + \vect{B}_{\vls,C}$ for BEC_$A$.
Aligning the magnetic bias field parallel to the split beam ($\mathbf{B}_0 \parallel \uvect{z}'$ in \reffig{fig:vls_super_fig}) ensures maximum sensitivity to the VLS field difference between beams $A$ and $B$, whilst minimizing any contribution to the differential signal from beam $C$.

As per Eqs.~(\ref{eq:Evls_defn}) and (\ref{eq:delta_B_approx}), the VLS field difference depends linearly on the intensity difference sampled by each BEC, and the common polarization circularity:
\begin{equation}
\label{eq:DeltaBvls}
    \Delta B_\vls = \frac{2\pi}{\gamma} \alpha_V \Delta I \sin 2\theta \,.
\end{equation}
The gradient of $(\Delta I, \Delta B_\vls)$ data yields the vector light shift per unit intensity, which in the vicinity of vanishing polarization circularity ($\abs{\theta} \ll 1$), is given by
\begin{equation}
\label{eq:dBvlsdI}
    \partialD{B_\vls}{I} = \pm \frac{4\pi}{\gamma} \alpha_V \left( \thetaQWP - \thetaN \right) \, .
\end{equation}
We have taken $\theta = \thetaQWP - \thetaN$, with $\thetaQWP$ the rotation angle of a quarter-wave plate used to control the circularity, offset by the `nulling angle' $\thetaN$ at which (a) $\theta = \mathcal{C} = 0$, and (b) the vector light shift is independent of intensity and is thus vanishing.
Minimizing the vector light shift therefore amounts to reducing the gradient $\partial B_\vls/\partial I$, i.e. by finding the nulling angle $\thetaN$.

\subsection{Controlling relative intensity}
We vary the VLS field difference by changing the intensity of one of the parallel dipole beams, $A$ or $B$.
For the same optical power in beams $A$ and $B$, the two traps they form are not identical; the beams have slightly different waists due to aberrations, they intersect beam $C$ at different axial and radial positions, etc. 
We define the intensity experienced by each condensate as $I_A(\vect{r}_A)$ and $I_B(\vect{r}_B)$.
Despite small differences in the intensity profile of each beam, and the relative location of the BECs in each beam (e.g. the vertical position of each BEC with respect to its respective beam center), the relevant independent variable is the intensity difference $\Delta I \equiv I_A(\vect{r}_A) - I_B(\vect{r}_B)$.
Operationally, this intensity difference is controlled by changing the rf power in one or both of the tones delivered to the AOM used to generate beams $A$ and $B$.
Denoting these rf powers by $P_A$ and $P_B$, and assuming that the intensity experienced by each BEC is linear in these powers, we have
\begin{subequations}
\begin{align}
    I_A(\vect{r}_A) &= p_A P_A \\
    I_B(\vect{r}_B) &= p_B P_B \, .
\end{align}
\end{subequations}
Supposing we vary the differential intensity by adjusting $P_A \mapsto P_A + \Delta P_A$, we have $I_A \mapsto I_A + \Delta I_A$, and
\begin{equation}
    \Delta I = p_A (P_A + \Delta P) - p_B P_B \,.
\end{equation}
The intensity experienced by each BEC is equal when $\Delta I = 0$, requiring a change in rf power modulating beam $A$ by an amount
\begin{equation}
    \Delta P_0 = \frac{p_B}{p_A} P_B - P_A \, .
\end{equation}
The differential power $\Delta P_0$ is determined experimentally by adjusting $\Delta P$ until the VLS-induced relative phase of the two condensates is independent of the polarization, i.e. the horizontal coordinate where lines of best fit to $(\Delta \phi$, $\Delta P)$ data intersect for varying beam polarizations.
We present such graphs using an offset rf power difference $\Delta P' \equiv \Delta P - \Delta P_0$, such that $\Delta P' = 0 \Leftrightarrow \Delta I = 0$.
The differential intensity can be conveniently expressed in terms of this offset rf power difference:
\begin{align}
    \Delta I &= p_A (P_A + \Delta P' + \Delta P_0) - p_A P_B \notag \\
             &= p_A \Delta P' \, , 
\end{align}
from which one can also equate a normalized optical intensity difference with a normalized rf power difference:
\begin{equation}
    \frac{\Delta I}{I_A} = \frac{\Delta P'}{P_A} \, .
\end{equation}
For the experiments described here, the normalized power difference required to balance the intensities is $\sim 33\%$.

Finally, expressing $\Delta I$ in absolute units of intensity requires knowledge of $p_A$, which can be determined from $(P_A, I_A(\vect{r}_A))$ data.
In practice, this calibration is performed by measuring optical power as a function of rf power, and inferring intensity from the measured beam profile.
The intensity sampled by both condensates when $\Delta I = 0$ is $I_A(\vect{r}_A) = I_B(\vect{r}_B) = \unit[8.39\times10^3]{W\,cm^{-2}}$.
This is independently verified by comparing the predicted intensity profile at the location of the condensate with measurements of the trapping frequencies of the dipole potential.

\subsection{In-trap interferometry results}
Background magnetic field gradients limited the maximum permissible interrogation time to $T \lesssim 2\pi/(2 r_{\text{TF}} \gamma B') = \unit[25]{ms}$ for a Thomas-Fermi radius $r_{\text{TF}} = \unit[13]{\upmu m}$ and an ambient gradient of $B' = \unit[22]{mG/cm}$, which we did not cancel for this experiment.
The quadratic Zeeman shift further modulates the transverse spin length and thus the interferometric contrast at angular frequency $2 q_Z$, with $q_Z = 2\pi \times \unit[71.89]{Hz\,G^{-2}}$ for \Rb.
We chose $T=\unit[15]{ms}$ to coincide with a local maximum in this contrast, which was as high as $80\%$.

\begin{figure}
    \centering
    \includegraphics[width=\columnwidth]{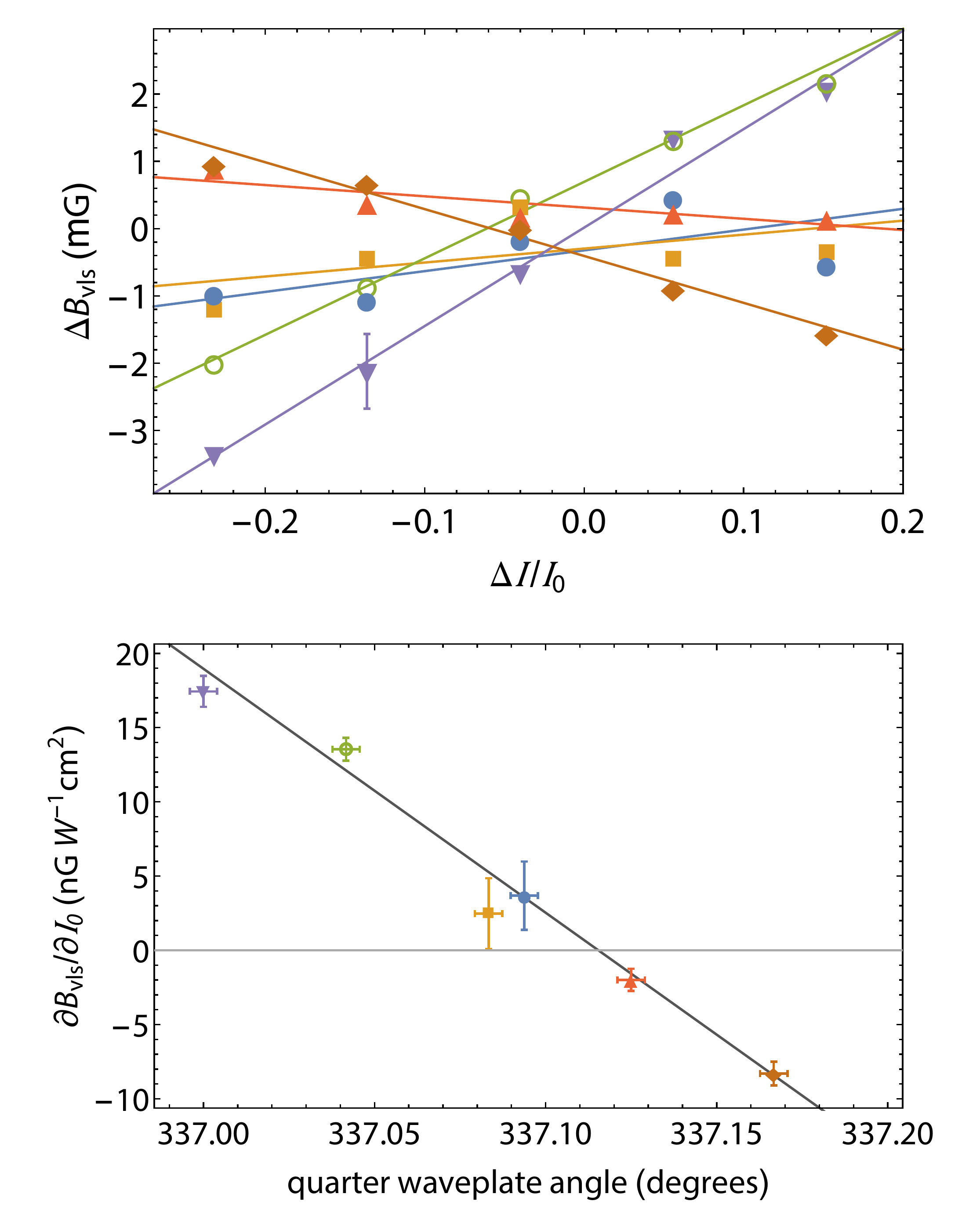}
    \caption{In-trap measurement and cancellation of VLS field difference. 
    (Top) The variation of the VLS field difference $\Delta B_\vls$ with intensity difference, for six quarter-wave plate angles over a $\unit[10]{arcmin}$ range.
    The single error bar is representative of typical uncertainty reported from ellipse fits, $\text{u}(\Delta \phi) \approx 0.011\,\pi$.
    The gradient of each line is an intensity normalized VLS field, as shown vs. waveplate angle in the lower plot.
    Vertical error bars represent the uncertainty in each fitted gradient $\partial B_\vls / \partial I$; horizontal error bars represent the resolution of the QWP rotation stage.
    The slope of this line can be used to impute the vector polarizability (\refeq{eq:dBvlsdI}), and a suppression ratio of $2.1(8)\times10^4$ (Eqs.~\ref{eq:gradient_to_amplitude} and \ref{eq:suppression_ratio}) corresponding to the minimum achieved intensity normalized VLS field.}
    \label{fig:VLS15ms}
\end{figure}

\reffig{fig:VLS15ms}(top) shows the differential VLS field -- created by beams $A$ and $B$ -- as a function of the normalized relative intensity at each condensate location, for six QWP angles $\thetaQWP$ in the vicinity of the nulling angle $\thetaN$.
Plotting the gradient of these data versus the QWP angle (\reffig{fig:VLS15ms}(bottom)) allows us to impute the parameters of \refeq{eq:dBvlsdI} from a line of best fit, namely the vector polarizability $\alpha_V$ and the nulling angle $\thetaN$.
From the slope of the line in the lower panel of \reffig{fig:VLS15ms}, we impute the reduced vector polarizability of \Rb ($F=1$) at $\lambda = \unit[1064]{nm}$ to be $\alpha^v_{nJF}=\unit[2.33(13)\times10^{-40}]{C\,m^2/V}$, in remarkable agreement with the theoretical value calculated in Section \ref{sec:theory}.
In experimentally-motivated units this corresponds to a reduced vector light shift per unit intensity $\alpha^v_{nJF}/(4 c \epsilon_0 F) = h \times 0.331(18)\,\HzWcm$ (cf. \refeq{eq:Evls_defn}).

The minimum intensity normalized VLS field we achieved was $-\unit[1.99(75)]{nG \,W^{-1}\,cm^2}$ for a QWP angle of $\thetaQWP=337.125(4)^\circ$.
To estimate the suppression of the VLS induced field, relative to its maximum value, we note that the maximum intensity normalized VLS (for $\abs{\mathcal{C}} = 1$) is half of the maximum gradient with respect to $\theta$ (for $\theta = 0$) owing to the form of \refeq{eq:DeltaBvls}, i.e.
\begin{equation}
\label{eq:gradient_to_amplitude}
    \left( \partialD{B_\vls}{I} \right)_{\text{max}} = \frac{1}{2} \left| \partialD{}{\theta} \partialD{B_\vls}{I} \right|_{\theta = 0} \, .
\end{equation}
The gradient with respect to $\theta$ is the slope of the fit in \reffig{fig:VLS15ms}(bottom), appropriately scaled for $\theta$ in radians.
Using this slope, the VLS suppression we achieved is described by the ratio
\begin{align}
\label{eq:suppression_ratio}
    \frac{
        \left(
            \partial B_\vls / \partial I 
        \right)_{\text{min, achieved}}
    }
    {
        \left(
            \partial B_\vls / \partial I
        \right)_{\text{max}}} 
    &= 2.1(8) \times 10^{-4} \, .
\end{align}
Alternatively, the suppression can be quantified by estimating how close the QWP could be controlled to the imputed nulling angle ($\thetaN = 337.115(3)^\circ$), i.e. \mbox{$2|\thetaQWP - \thetaN|_{\text{min, achieved}} = 3.3(1) \times 10^{-4}$}.
In either case, the suppression ratio is independent of any systematics associated with the intensity calibration.

While small, the residual temperature dependence of the retardance of the zero-order waveplates is likely to be the dominant source of statistical uncertainity.
The laboratory temperature is controlled to better than \unit[0.25]{K} accuracy and rms fluctuations over a typical experimental sequence are usually below \unit[0.1]{K}.
The nulling experiment required a considerable number of BEC shots and ran over several hours, much longer than the characteristic times of temperature control in the laboratory.
We thus consider that the ordinate error bars fully account for temperature 'drift' in the laboratory. 

Two main systematics limit our ability to comprehensively null the VLS: the resolution of the QWP rotation stage ($\unit[0.1]{mrad}$), and the presence of increased statistical noise in the measured gradients around the nulling angle.
The former may be circumvented by using higher resolution rotation stages, which would additionally allow for longer interrogation times that amplify the slope of \reffig{fig:VLS15ms}.
Longer interrogation times are possible provided loss of interferometric contrast due to background field gradients is suppressed.

The origin of increased variation of $\partial B_\vls / \partial I$ in the vicinity of the nulling angle is not immediately apparent.
The \emph{relative} polarization difference between the two dipole beams $A$ and $B$ is assumed to be small, as both beams begin with a well-defined linear polarization and travel similar optical paths through the same components.
However, we cannot discount the possibility that a relative polarization imperfection exists, due to spatially varying birefringence of the glass vacuum cell or other components, such as the QWP itself.
Any polarization difference between the two beams would result in a relative VLS that is impossible to cancel.

We also performed measurements to quantify the contribution to the measured VLS field difference from the crossing beam, which should be minimal due to the bias field alignment along $z'$.
Rotating the angle of the QWP in the path of beam $C$ by $10^{\circ}$, we observed no significant excursion outside of measured uncertainties (cf. the entire range of data plotted in \reffig{fig:VLS15ms} is $\unit[10]{arcmin}$).

Accurately quantifying the atomic vector polarizability relies on a sound measure of the intensity at the atoms, a problem that vexes other measurements of atomic polarizabilities from AC Stark measurements~\cite{mitroy_theory_2010}.
The local intensity of a dipole trapping beam sampled by the trapped cloud is determined by the position of the trap minimum, which is gravitationally shifted from the point of peak intensity.
The addition of the crossing beam further complicates the problem, rendering any intensity characterization without a direct atomic metric of limited applicability.
A precise, spatially resolved measurement of absolute optical intensity with an atomic measurement is a significant experimental achievement in itself.
This uncertainty in the absolute intensity sampled by the two condensates results in a systematic error in our estimate for $\alpha_V$ at $\unit[1064]{nm}$.
These issues in no way detract from the utility of our technique for nulling the VLS induced field from the trapping beams.

\section{Coherent spin-mixing}
\label{sec:spin_mixing}
\begin{figure*}
    \centering
    \includegraphics[width=\textwidth]{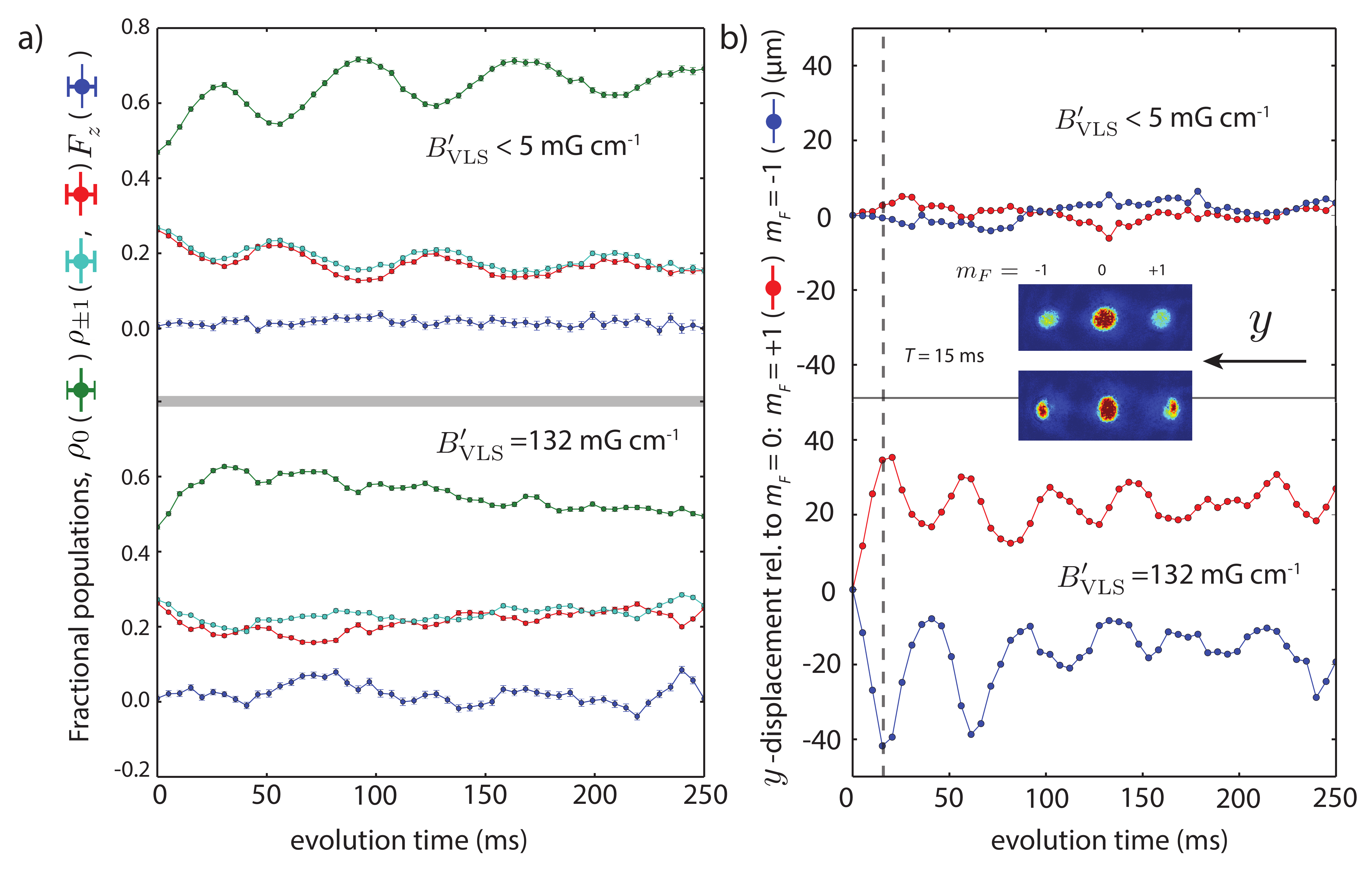}
    \caption{
      \label{fig:spin_mixing_composite}
        Spin-mixing dynamics of a spinor BEC with the VLS approximately nulled (top) and when set to $\unit[132(6)]{mG\,cm^{-1}}$ (bottom): fractional population evolution (left) and centroid displacement, relative to the $\rho_0$ centroid position  (right).
        The VLS induces an effective magnetic field gradient along the gravitational direction ($y$), driving strong component separation (shown in the inset Stern-Gerlach absorption image, taken $\unit[15]{ms}$ after the $\pi/2$ pulse). 
        Separation reduces spatial overlap of the three Zeeman sublevels and suppresses spin mixing from the $m_F = \pm1$ states into the $m_F = 0$ state.
    } 
\end{figure*}
To demonstrate the utility of VLS control, we consider the spin mixing dynamics of a spinor BEC in the presence of a trap induced VLS.
Coherent spin mixing is one of the most interesting facets of spinor BECs~\cite{chang_coherent_2005, kawaguchi_spinor_2012, stamper-kurn_spinor_2013}, vividly demonstrating the unique nature of spinor quantum fluids and with emerging application to quantum enhanced sensing~\cite{linnemann_quantum-enhanced_2016}. 
Coherent spin mixing collisions transfer population between the Zeeman states at low magnetic field: a colliding pair of atoms in the $m_F = 0$ state can create an $m_F = \pm1$ pair, and vice-versa:
\begin{equation}
\label{eq:spin_mixing}
    \ket{m_F = +1} + \ket{m_F = -1}\rightleftharpoons \ket{m_F = 0 } + \ket{m_F = 0}.
\end{equation}
The energy difference between the two pairs is dictated by the quadratic Zeeman shift and the spin-dependent collisional interaction energy, and is of order $\unit[10]{Hz}$ for the $F=1$ levels of $^{87}$Rb and order $\unit[100]{Hz}$ for $^{23}$Na at typical bias fields of order $0.1$--$\unit[1]{G}$.
This energy is typically smaller than the chemical potential, and consequently the spin-changing two-body interaction manifests as coherent oscillations in the Zeeman populations while the atom density retains a single spatial mode.
While an overall linear Zeeman shift plays no role in these dynamics, even a weak magnetic field \emph{gradient} of order~$\unit[1]{mG/cm}$ is sufficient to create a linear Zeeman shift gradient across the condensate exceeding the collisional and quadratic Zeeman energy scales, invalidating the single mode approximation~\cite{zhang_magnetization_2011}.

Spin mixing is initiated by putting the condensate into a superposition different to the mean-field spinor ground state.
A single $\pi/2$-pulse is applied to a condensate in the $m_F = -1$ state.
Denoting the relative population fractions in each Zeeman sublevel by $\rho_{m_F} = N_{m_F}/N$, the initial populations immediately after the rf pulse are $(\rho_{-1}, \rho_{0}, \rho_{+1}) = (0.25,0.5,0.25)$.
Spin-mixing collisions conserve the overall magnetization, $m = \rho_{+1} - \rho_{-1} = \langle F_z\rangle$, leading to oscillation of the population in each Zeeman sublevel.

In the presence of a gradient magnetic field, mechanical forces separate the magnetically sensitive $m_F = \pm1$ components, leading to dephasing, and in the case of a Ramsey experiment, loss of interferometric contrast.
This is reflected in the spin population dynamics as a damping of the amplitude of the oscillations: component separation results in less spatial overlap of the Zeeman components.
The reduction in oscillation amplitude is indicative of a weakening of the spin interaction strength.
When attributable to gradients  it is undesirable in experiments observing long-term spinor evolution~\cite{liu_number_2009, guzman_long-time-scale_2011}, where weak equilibration mechanisms may have similar effects.

A bias field of $\unit[372]{mG}$ parallel to the $x'$-oriented dipole beam was applied, resulting in maximum sensitivity to the VLS from that dipole beam.
The magnitude of the quadratic Zeeman shift is $q = 2\pi \times \unit[10]{Hz}$, and the spinor interaction energy is given by $c = c_2 \langle n \rangle$, with $c_2 = -\unit[2.4\times10^{-53}]{J\,m^3}$ for \Rb and $\langle n \rangle$ the average number density: in our experiments, $c \approx - 2\pi \times \unit[3.2]{Hz}$.
The background magnetic field gradient in this experiment was $\partial B/\partial z' < \unit[6]{mG\,cm^{-1}}$.

\reffig{fig:spin_mixing_composite} shows the population dynamics of a spinor condensate in the optical dipole trap, with the VLS approximately nulled (top) and when set to maximum (bottom).
Population oscillations are barely discernible in the lower panel taken at full VLS, and the images show substantial component separation along the gravitational direction (as expected from the sagging dipole potential), indicated by the centroid motion shown to the right in \reffig{fig:spin_mixing_composite}.
The reduced spatial overlap of the $m_F = \pm1$ components prevents spin mixing into the $m_F=0$ state, and the $\rho_0$ population is gradually exhausted as the production of $m_F = \pm1$ pairs is favored~\cite{stamper-kurn_spinor_2001}.
With the VLS nulled to below $\unit[5]{mG\,cm^{-1}}$, more than three periods of population oscillations were clearly visible, and component separation was suppressed by an order of magnitude.
An underlying gradual increase in $\rho_0$ population was highly reproducible, and we suspect this relaxation behavior was a result of the background \emph{magnetic} field gradient (as opposed to the `fictitious' gradient induced by the VLS of the trap beam).
This behavior resembles the longer-timescale relaxation observed previously~\cite{guzman_long-time-scale_2011}.
Further study of the dynamics of spinor BECs in well-characterized gradients is an interesting prospect, with optical potentials of specially-tailored VLS are used to engineer specific Zeeman phases on a spinor BEC, and observe other gradient-induced effects, such as non-conservation of magnetization~\cite{zhang_magnetization_2011}.

\section{Conclusions}
We have demonstrated high-precision interferometric measurement of the vector light shift from a crossed beam optical dipole trap.
Using this measurement, we determined the VLS as a function of intensity, and hence optimized polarization until the VLS was eliminated.
We measured the VLS contribution from each beam separately; so that the residual VLS was readily nulled by separate adjustments of waveplates in each beam.
Removing the VLS removed the VLS-induced effective magnetic field gradient across our spinor condensate.
Nulling these gradients allowed us to observe slow spin-mixing dynamics in the spinor BEC, previously obscured by Stern-Gerlach component separation in the VLS-induced gradient.
While our measurements specifically considered the case of eradicating residual spurious shifts from an optical dipole trap, our techniques may have interesting application in the characterization of VLS shifts from other optical fields, for the purposes of tailoring microscale Zeeman-shift potentials far more intricate than those attainable with magnetic fields.  
Our results will be of interest to experiments investigating the fundamental physics and metrological applications of trapped spinor gases.

\begin{acknowledgments}
This work was supported by the Australian Research Council (DP1094399) and the Australian Postgraduate Award Scheme.
\end{acknowledgments}

\appendix*

\section{Thermally-induced birefringence in cell windows and trap optics}
Changes in birefringence over the duration of an experimental shot or a contiguous sequence of shots are not nulled by our method.
The primary cause of drifting birefringence is varying temperatures of elements designed to be birefringent (such as waveplates) or with long optical paths (such as optical fibers).
The solutions are well-known: control of laboratory temperature, use of zero-order waveplates, appropriately placed efficient (crystal) polarizers to `clean up' polarization linearity (at the expense of small intensity drift), use and careful alignment of polarization-maintaining fibers etc.

Once these obvious sources of birefringence are controlled, the question arises as to whether the vacuum cell windows themselves contribute birefringence.
Vacuum cells are typically made of isotropic materials such as fused silica which have low, but not zero, birefringence due to residual stress remaining in the material from the manufacturing process; this is static and is nulled in our process.
However stresses on the cell caused by heating due to dipole beam absorption are neither static nor easily avoided.
Conventional wisdom is that such thermally-induced birefringence is negligible in highly-transparent isotropic materials such as fused silica.

We provide an educated estimate here confirming that dipole-beam induced thermal stress birefringence is small, and is comparable to the sensitivity limit we have reported here only if the cell is exposed to especially intense and prolonged trap beam irradiation. 
Estimating the space- and time-varying birefringence of the cell window requires first finding the optical power absorbed in the material, then the resultant temperature rise, the thermal stress this causes in a fixed material, and finally the consequent stress-induced birefringence.
We use fused silica as an example as it is a common cell material with generally well-specified optical, thermal, elastic and photoelastic properties~\cite{schott_lithosil_2006}.
For high-power interferometer applications of fused silica elements the depolarizing effects of thermal birefringence were found to be much less significant than thermal lensing effects~\cite{winkler_birefringence-induced_1994}. 
However in atomic magnetometry applications the emphasis is on VLS-inducing birefringence much more than wavefront distortion. 

The least well-specified property for fused silica is the volume attenuation coefficient, which is sufficiently small that measurement is challenging for reasonable thicknesses of glass and in the presence of Fresnel loss~\cite{yoshida_method_2003}.
These measurements reported $\mu = \unit[0.0018(4)]{m^{-1}}$, and ultrapure substrates prepared for LIGO have reportedly achieved $\mu \approx \unit[10^{-4}]{m^{-1}}$~\cite{strain_thermal_1994}.
Notwithstanding these measurements, we adopt the Schott specification of at least 99.9\% transmission for 10\,mm thick fused silica for wavelengths longer than \unit[250]{nm}, giving an absorption coefficient of $\mu < \unit[0.1]{m^{-1}}$, and providing a conservative upper bound for absorption across the full range of wavelengths in use for optical trapping.
Assuming a constant dipole trap power of $P_\text{ODT}=\unit[10]{W}$ and a cell thickness of $d=\unit[5]{mm}$, the absorbed power is at most $P\approx \mu d P_\text{ODT}=\unit[5]{mW}$.

Absorbed power is deposited across the Gaussian beam profile; for our nominal trap waist of $\unit[80]{\upmu m}$ with cell windows surfaces \unit[25]{mm} and \unit[30]{mm} from the trap, the cell is less than 2 Rayleigh ranges from the waist and the beam radius ranges from $\unit[130]{\upmu m}$ to $\unit[160]{\upmu m}$ through the cell.
Few dipole traps will have beam radii at the cell much smaller than this: much tighter waists will have larger beam diameters at the cell, and large volume ODTs with larger waists will include the cell within a Rayleigh range; the exceptions are traps located very close to the inner window surface.

The temperature rise is found by solving the heat equation for an infinite medium with a constant power deposition per unit length across a Gaussian profile through it.
A closed-form solution is available~\cite{antonakakis_closed_2013}, but it is sufficient for our estimate to consider the temperature rise on axis,
given by
\begin{equation}
  \Delta T(t) = T_0 \ln \left(1+\frac{2 D t}{w^2}\right).
  \label{eq:logtemp}
\end{equation}
where $T_0 = P_\text{abs} / 4\pi k d \approx \unit[60]{mK}$ is the characteristic temperature with $k=\unit[1.31]{W\,m^{-1}\,K^{-1}}$ the thermal conductivity, $w$ is the mean beam radius through the cell, and $D$ is the thermal diffusivity.

At short times after the beam is turned on the window temperature inside the volume of the beam rises linearly, with heat contained in the beam volume. 
The thermal diffusivity of fused silica is $D=\unit[7.5\times 10^{-7}]{m^2/s}$~\cite{schott_lithosil_2006}, and so the time for heat to diffuse beyond beam radius $w=\unit[145]{\upmu m}$ is $\tau = w^2/D \approx \unit[30]{ms}$.
Once the diffusion length becomes comparable to the cell thickness the model of an infinitely thick medium becomes inappropriate; this occurs for times approaching $d^2/D \approx \unit[30]{s}$.
Thus for times between \unit[100]{ms} and \unit[10]{s}, relevant to most dipole trap experiments, the axial temperature rise should be reasonably well modeled by \refeq{eq:logtemp}, increasing only logarithmically with time to around $6.6T_0$.
For our assumed constant power dissipation per unit length of $P_\text{abs}/d=\unit[1]{W/m}$ the axial temperature rise is $\Delta T_\text{max} = \unit[400]{mK}$ after \unit[10]{s}. 

The temperature rise varies only radially inducing a plane-stress tensor distribution $\sigma_{ij}(r)$.
It can be shown that the stresses are maximal on axis, taking the principal values
\begin{equation}
  \sigma_{rr}(0) = \sigma_{\theta\theta}(0) = \frac{1}{2} \alpha_\text{CTE} E \Delta T_\text{max},
\end{equation}
where $E=\unit[72]{GPa}$ is the Young modulus, and $\alpha_\text{CTE}=\unit[5\times 10^{-7}]{K^{-1}}$ the coefficient of thermal expansion, of fused silica.
For our estimated values the radial and tangential principal stresses on axis are \unit[7.2]{kPa}.

It is then straight-forward, at least in an isotropic medium such as fused silica, to infer optical path difference due to stress birefringence $\text{OPD} = K d \Delta \sigma$, where $\Delta \sigma = \sigma_{rr} - \sigma_{\theta\theta}$ is the difference in stresses between the first and second principal axes~\cite[\S5.5]{laufer_introduction_1996}, and $K=\unit[3.4\times 10^{-12}]{Pa^{-1}}$ is the relative stress-optic coefficient of fused silica. 
On the axis $\sigma_{rr} = \sigma_{\theta\theta}$ and so the birefringence vanishes, as it must due to the rotational symmetry.
Nevertheless, we can place an upper bound on the stress birefringent path difference off-axis of $K \sigma_{rr} d \approx \unit[1.2\times 10^{-10}]{m}$ or a retardance of $\theta=2\pi\times \unit[1.2 \times 10^{-10}]{m}/\lambda = \unit[7 \times 10^{-4}]{rad}$.

A full thermoelastic analysis of a Gaussian-beam-heated medium~\cite{mosca_photon_2010,khazanov_compensation_2004} shows that the birefringent retardance in the $t \gg d^2/D$ limit is
\begin{equation}
  \theta(r) = \frac{4\pi d}{\lambda} Q T_0 \left (1+\frac{\exp(-2r^2/w^2)-1}{2r^2/w^2}\right),
\end{equation}
with $Q=n_0^3 \alpha_\text{CTE} (1+\nu)(p_{11}-p_{12})/4(1-\nu) = \unit[-8.0\times 10^{-8}]{K^{-1}}$ termed the effective optoelastic coefficient~\cite{mosca_photon_2010}, $n_0 = 1.45$ the cold refractive index, $\nu=0.17$ Poisson's ratio, and $p_{11}=0.121,p_{12}=0.270$~\cite{winkler_birefringence-induced_1994} the only non-vanishing components of the photoelastic tensor for isotropic fused silica.
Thus despite the overall temperature of the window continuing to increase, the retardance approaches a constant value far off-axis of $\theta_\text{max} = 4 \pi d Q T_0/\lambda \approx \unit[1.4\times 10^{-4}]{rad}$, and $\unit[8\times 10^{-5}]{rad}$ when $r=w$, i.e. one $1/e^2$ radius of the beam off-axis, less than one order smaller than our upper bound estimate.

For a perfectly linearly polarized beam incident on the cell window, this radial retardance pattern induces a azimuthally-varying polarization~\cite{mosca_photon_2010,amjad_laser-induced_2011} of peak circularity $\mathcal{C}=\sin 2 \theta \approx 3\times 10^{-4}$, which is just below the limit of detectability with the differential Ramsey interferometer.
In practice the circularity measured by the atoms will be even smaller for several reasons: some experiments will measure only the net circularity of the beam which may be much smaller again; the above parameters are highly conservative and assume a very intense dipole trap laser heating the glass for a relatively long period; most significantly, fused silica may be orders of magnitude more transparent than the specification we have adopted here.
Optical dipole traps typically vary intensity on time scales slow compared to elastic timescales (microseconds) but comparable to thermal timescales (millisecond to seconds); a more careful model demands a full spatiotemporal thermal analysis~\cite{farrukh_analysis_1988} feeding the quasistationary photoelastic analysis~\cite{khazanov_compensation_2004,mezenov_thermooptics_1987}.

At all other optical elements transmitted by the dipole trap beam, the beam radius is an order of magnitude larger, and the thermal diffusion time thus two orders longer: surface heat loss will ensure the temperature rise is much lower than in the cell windows.
On first consideration the final waveplates before the cell may also be suspect for thermal stress birefringence, as these elements are strongly birefringent.
However the (many) photoelastic tensor components of $\alpha$-quartz~\cite{narasimhamurty_photoelastic_1969} are of the same order as fused silica, and the much shorter optical path, part cancellation of the stress birefringence by the fast/slow lamination, and much shallower temperature gradients make stress birefringence in the waveplates genuinely negligible.
Longitudinal thermal expansion in zero-order waveplates decreases retardance by order $\unit[10^{-4}]{rad/K}$, and so can be neglected for expected temperature rises much less than \unit[1]{K}.

In conclusion, the great transparency and very low thermal coefficient of expansion of fused silica more than compensate for its somewhat higher stress optic coefficient compared to most glasses, and render fused silica only very weakly birefringent under the thermal stress of intense laser illumination.
However other configurations, notably dipole traps used to hold single atoms close to cell surfaces, may have intensities orders of magnitude higher and thermal stress birefringence must be considered carefully in such cases.

\bibliography{vls_ramsey}
   
\end{document}